# GEDI: Scalable Algorithms for Genotype Error Detection and Imputation

Justin Kennedy*    Ion I. Măndoiu*    Bogdan Paşaniuc†

**Abstract**

Genome-wide association studies generate very large datasets that require scalable analysis algorithms. In this report we describe the GEDI software package, which implements efficient algorithms for performing several common tasks in the analysis of population genotype data, including genotype error detection and correction, imputation of both randomly missing and untyped genotypes, and genotype phasing. Experimental results show that GEDI achieves high accuracy with a runtime scaling linearly with the number of markers and samples. The open source C++ code of GEDI, released under the GNU General Public License, is available for download at http://dna.engr.uconn.edu/software/GEDI/.

## 1 Introduction

Genome-wide association studies (GWAS) have recently led to the discovery of hundreds of genes associated with complex human diseases [6]. Each such study involves genotyping thousands of cases and controls at hundreds of thousands of single nucleotide polymorphism (SNP) loci, generating very large datasets that require scalable analysis algorithms.

In this report we decribe GEDI, a software package implementing efficient algorithms for several common tasks in the analysis of GWAS data:

- **Genotype error detection and correction.** Despite continuous progress in technology and calling algorithms, errors remain present in high-throughput SNP genotype data [16] at levels that can invalidate statistical test for disease association, particularly for haplotype-based methods [10]. GEDI implements the likelihood ratio approach to error detection from [9].

- **Missing data recovery.** High-throughput genotyping platforms leave uncalled large numbers of SNP genotypes. To complement quality control procedures that exclude SNP loci and samples with high proportions of missing genotypes, GEDI provides methods for maximum-likelihood inference of remaining missing genotypes.

---

*Computer Science & Engineering Department, University of Connecticut, Storrs, CT 06269, USA
†International Computer Science Institute, Berkeley, CA 94704, USA

- **Imputation of genotypes at untyped SNP loci.** Current genotyping platforms allow simultaneous typing of as many as a million SNP loci, but this is still just a fraction of the polymorphisms present in the human population. Imputation of genotypes at untyped SNP loci based on linkage disequilibrium information extracted from reference panels such as Hapmap [8] is often performed to increase statistical power of GWAS studies, see, e.g., [13]. Furthermore, imputation is critical for performing meta-analysis of datasets generated using different platforms [17].

- **Genotype phasing.** Haplotype based association tests can improve statistical power compared to single-SNP approaches, but have seen limited use in the analysis of GWAS data, in part due to the lack of haplotype inference methods that are both accurate and scalable. In an attempt to fill in this gap, GEDI includes an implementation of the highly-scalable phasing algorithm of [5], based on entropy minimization. This algorithm has been recently used by [1] in conjunction with a haplotyping sharing approach to implicate in Parkinson's disease a novel gene missed by traditional single-SNP analyses.

## 2 Statistical Model

At the core of GEDI's algorithms is a factorial hidden Markov model (HMM) of multilocus SNP genotype data (Fig. 1). Under this model, a multilocus genotype is formed by random pairing of haplotypes obtained as the result of historical recombination among $K$ founder haplotypes, where $K$ is a user-selected parameter. More specifically, the founder haplotypes of origin for SNP alleles on each autosomal chromosome are assumed to form a first order HMM whose parameters are estimated using the Baum-Welch algorithm based on reference haplotypes (for imputation of untyped SNP genotypes) or a pool consisting of reference haplotypes and haplotypes inferred from the genotype data itself (for error detection and missing data recovery).

Formally, we denote by 0 and 1 the major and minor alleles at every biallelic SNP locus, by 0, 1, and 2 the three possible SNP genotypes (homozygous major/minor, respectively heterozygous), and by '?' a missing SNP genotype. Multilocus SNP genotypes are modeled statistically using the factorial HMM represented graphically in Fig. 1. Formally, for every SNP locus $i \in \{1, \ldots, n\}$, we let $H_i, H_i' \in \{0, 1\}$ be random variables representing the alleles observed at this locus on the maternal (paternal) chromosome of the individual under study, and $F_i, F_i' \in \{1, \ldots, K\}$ be random variables denoting the founder haplotype from which $H_i$ and $H_i'$ originate, respectively. Values taken by these random variables are denoted by the corresponding lowercase letters (e.g., $g_i$, $h_i$, $f_i$, etc.). We set $P_M(g_i|h_i, h_i')$ to 1 if $g_i = h_i + h_i'$ and 0 otherwise. Probabilities $P_M(f_1)$, $P_M(f_{i+1}|f_i)$, and $P_M(h_i|f_i)$ are estimated using the classical Baum-Welch algorithm based on haplotypes inferred from a panel representing individual's population of origin, while $P_M(f_1')$, $P_M(f_{i+1}'|f_i')$, and $P_M(h_i'|f_i')$ are obtained by simply duplicating these estimates. We emphasize that GEDI estimates all HMM parameters based on the available data, without making any assumptions regarding the population under study. This underscores the robustness of its statistical model and its applicability to any studied population.

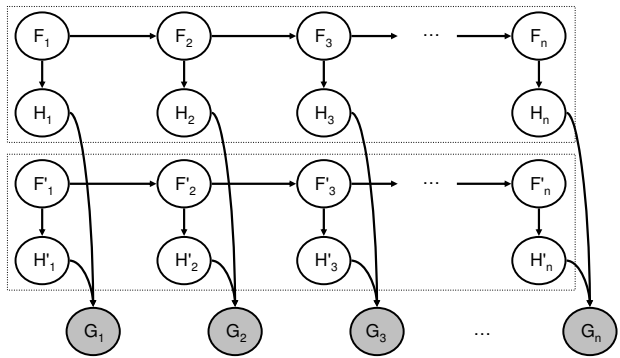

Figure 1: GEDI's factorial HMM model for multilocus SNP genotype data.

We denote by $\mathbf{g}[g_i \leftarrow x]$ the multilocus genotype obtained from $\mathbf{g} = (g_1, \ldots, g_n)$ by setting the value of the (possibly missing) $i$-th SNP genotype to $x$. For a trained factorial HMM $M$, GEDI uses a forward-backward algorithm (see next section for details) to efficiently compute all probabilities $P_M(\mathbf{g}[g_i \leftarrow x])$. GEDI performs error detection by computing the likelihood ratio $\max_x P_M(\mathbf{g}[g_i \leftarrow x])/P_M(\mathbf{g})$ for each non-missing SNP genotype $g_i$, and flagging a genotype as a potential error whenever this ratio exceeds a user specified threshold. A missing SNP genotype at a typed SNP locus $i$ is replaced by $\mathrm{argmax}_x P_M(\mathbf{g}[g_i \leftarrow x])$. Imputation of genotypes at an untyped SNP locus is performed similarly, except that in this case genotype probabilities are computed based on a "local" HMM model that spans the untyped locus and a user-specified number of typed SNP loci flanking it on each side.

## 3 Algorithmic and Implementation Details

GEDI computes multilocus genotype probabilities under a trained factorial HMM $M$ using an efficient forward-backward algorithm. Let $\mathbf{g} = (g_1, \ldots, g_n) \in \{0, 1, 2, ?\}^n$ be a given multilocus genotype. For every $i = 1, \ldots, n$, we define the forward and backward probabilities as $\mathcal{F}^i_{f_i, f_i'} = P_M(g_1, \ldots, g_{i-1}, f_i, f_i')$ and $\mathcal{B}^i_{f_i, f_i'} = P_M(g_{i+1}, \ldots, g_n | f_i, f_i')$, respectively. Then $P_M(\mathbf{g}[g_i \leftarrow x]) = \sum_{f_i=1}^K \sum_{f_i'=1}^K \mathcal{F}^i_{f_i, f_i'} \mathcal{B}^i_{f_i, f_i'} \mathcal{E}^i_{f_i, f_i'}(x)$, where $\mathcal{E}^i_{f_i, f_i'}(x) = \sum_{h_i + h_i' = x} P_M(h_i|f_i) P_M(h_i'|f_i')$. Thus all probabilities $P_M(\mathbf{g}[g_i \leftarrow x])$, $i = 1, \ldots, n$, $x = 0, 1, 2$, can be computed in $O(nK^2)$ once the forward and backward probabilities $\mathcal{F}^i_{f_i, f_i'}$ and $\mathcal{B}^i_{f_i, f_i'}$ are available.

The forward probabilities can be computed in $O(nK^4)$ using the recurrence:

$$\mathcal{F}^1_{f_1, f_1'} = P_M(f_1) P_M(f_1') \quad (1)$$

$$\mathcal{F}^i_{f_i, f_i'} = \sum_{f_{i-1}=1}^K \sum_{f_{i-1}'=1}^K \mathcal{F}^{i-1}_{f_{i-1}, f_{i-1}'} P_M(f_i|f_{i-1}) P_M(f_i'|f_{i-1}') \mathcal{E}^{i-1}_{f_{i-1}, f_{i-1}'}(g_{i-1})$$

$$= \sum_{f_{i-1}=1}^K P_M(f_i|f_{i-1}) \sum_{f_{i-1}'=1}^K \mathcal{F}^{i-1}_{f_{i-1}, f_{i-1}'} P_M(f_i'|f_{i-1}') \mathcal{E}^{i-1}_{f_{i-1}, f_{i-1}'}(g_{i-1}) \quad (2)$$

where $f_i, f_i' \in \{1, \ldots, K\}$, $i \in \{2, \ldots, n\}$. However, the inner sum in (2) is independent of

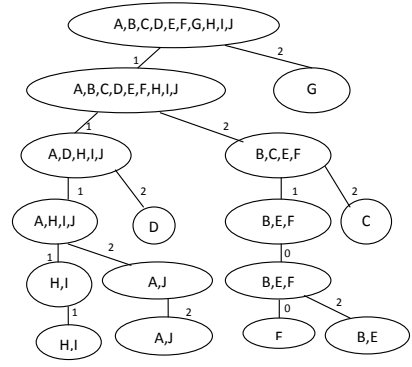

| Ind | SNP1 | SNP2 | SNP3 | SNP4 | SNP5 |
|-----|------|------|------|------|------|
| A | 1 | 1 | 1 | 2 | 2 |
| B | 1 | 2 | 1 | 0 | 2 |
| C | 1 | 2 | 2 | 0 | 2 |
| D | 1 | 1 | 2 | 2 | 2 |
| E | 1 | 2 | 1 | 0 | 2 |
| F | 1 | 2 | 1 | 0 | 0 |
| G | 2 | 1 | 2 | 1 | 1 |
| H | 1 | 1 | 1 | 1 | 1 |
| I | 1 | 1 | 1 | 1 | 1 |
| J | 1 | 1 | 1 | 2 | 2 |

(a)      (b)

Figure 2: Sample dataset over 5 SNPs (a) and corresponding trie (b).

$f_i$, and so its repeated computation can be avoided by replacing (2) with:

$$\mathcal{C}^i_{f_{i-1},f'_i} = \sum_{f'_{i-1}=1}^{K} \mathcal{F}^{i-1}_{f_{i-1},f'_{i-1}} P_M(f'_i|f'_{i-1}) \mathcal{E}^{i-1}_{f_{i-1},f'_{i-1}}(g_{i-1}) \qquad (3)$$

$$\mathcal{F}^i_{f_i,f'_i} = \sum_{f_{i-1}=1}^{K} P_M(f_i|f_{i-1}) \mathcal{C}^i_{f_{i-1},f'_i} \qquad (4)$$

By using recurrences (1), (3), and (4), all forward probabilities can be computed in $O(nK^3)$ time. Similarly, backward probabilities can be computed in $O(nK^3)$ time by using:

$$\mathcal{B}^n_{f_n,f'_n} = 1 \qquad (5)$$

$$\mathcal{D}^i_{f_{i+1},f'_i} = \sum_{f'_{i+1}=1}^{K} \mathcal{B}^{i+1}_{f_{i+1},f'_{i+1}} P_M(f'_{i+1}|f'_i) \mathcal{E}^{i+1}_{f_{i+1},f'_{i+1}}(g_{i+1}) \qquad (6)$$

$$\mathcal{B}^i_{f_i,f'_i} = \sum_{f_{i+1}=1}^{K} P_M(f_{i+1}|f_i) \mathcal{D}^i_{f_{i+1},f'_k} \qquad (7)$$

## 3.1 Trie Speed-up

For a dataset consisting of $m$ samples (i.e., multi-locus genotypes), running the forward-backward algorithm independently on each sample results in a runtime of $O(mnK^3)$, where $n$ is the number of SNP loci, and $K$ is the number of founder haplotypes. However, independent processing of the samples leads to repeated computation of forward and backward probabilities corresponding to genotype prefixes (respectively suffixes) shared by multiple genotypes. To avoid this, GEDI builds a prefix tree, or trie, from the given multilocus genotypes (see Fig. 2 for an example) and then computes probabilities by performing a preorder traversal of the trie. Computation of backward probabilities is sped-up in a similar way using a trie of reversed genotypes.

The speed-up achieved by using tries depends on the number and the similarity of the samples, as well as the number of SNP loci. For example, when performing imputation using 10 flanking SNPs on the 2502 samples of the IMAGE dataset described in next section, using tries gives a speed-up of 3×.

# 4 Experimental Results

For empirical evaluations of GEDI's error detection and phasing algorithms, and for a comparison of imputation algorithms implemented by GEDI and several other publicly available software packages including [2, 3, 11, 12, 13, 15] we direct the reader to [9], [5], respectively [4]. Here we present experimental results exploring the effect of GEDI's user-selected parameters on imputation accuracy.

Imputation experiments were performed on the Perlegen 600k genotype data (dbGaP accession number phs000016.v1.p1) generated by the International Multisite ADHD Genetics (IMAGE) project, comprising 958 parents-child trios from seven European countries and Israel. After excluding trios with one or more samples removed by data cleaning steps described in [14], we randomly selected 100 trios and phased them using the entropy minimization algorithm and pooled parental haplotypes with the 120 CEU haplotypes from Hapmap release 22 to form a reference panel of 520 haplotypes. The test data consisted of the genotypes of remaining 2502 IMAGE individuals, treated as unrelated unless otherwise indicated. Specifically, we masked 9% of the typed SNP loci on chromosome 22 (530 out of 5835), and computed the imputation error rate as the percentage of discordant imputations out of the total of 1,326,060 masked SNP genotypes. In all imputation experiments we used 10 typed SNP loci on each side of masked loci, which, as shown in Fig. 3, yields an excellent tradeoff between accuracy and runtime.

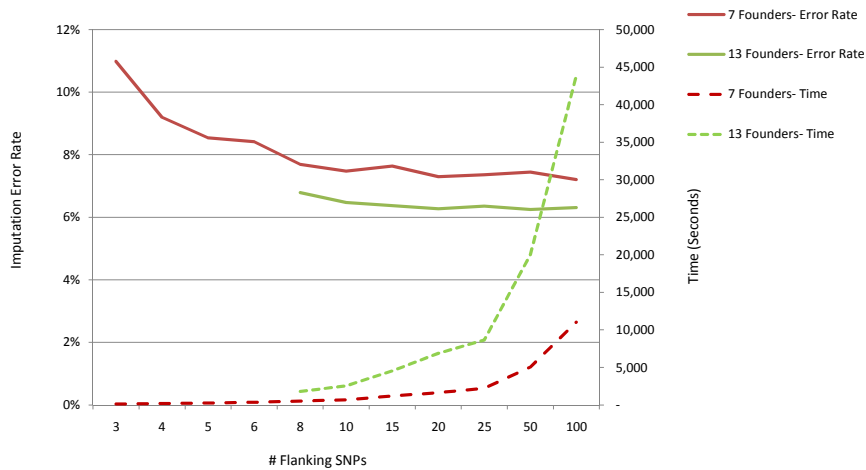

Figure 3: Imputation error rate and runtime for varying number of flanking typed SNP loci (IMAGE chr. 22 dataset, 520 training haplotypes).

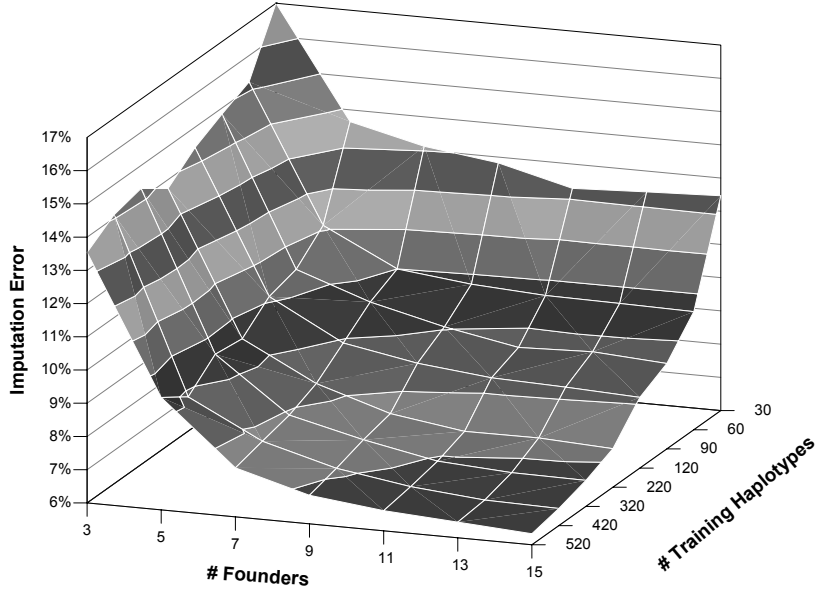

Figure 4: Imputation error rate on the IMAGE chr. 22 dataset for varying numbers of HMM founders and training haplotypes.

Table 1: Imputation error rate on the IMAGE chr. 22 dataset for varying numbers of HMM founders and training haplotypes.

| # Training Haplotypes | # Founders | | | | | | |
|---|---|---|---|---|---|---|---|
| | 3 | 5 | 7 | 9 | 11 | 13 | 15 |
| 30 | 17.02% | 13.65% | 13.11% | 12.82% | 12.27% | 12.37% | 12.47% |
| 60 | 15.21% | 11.10% | 10.00% | 9.75% | 9.62% | 9.59% | 9.55% |
| 90 | 14.82% | 10.35% | 9.58% | 9.04% | 8.63% | 8.71% | 8.57% |
| 120 | 14.39% | 10.11% | 8.93% | 8.52% | 8.30% | 8.23% | 8.13% |
| 220 | 13.73% | 9.42% | 8.28% | 7.58% | 7.26% | 7.27% | 7.16% |
| 320 | 14.31% | 9.53% | 7.91% | 7.37% | 6.94% | 6.81% | 6.78% |
| 420 | 14.10% | 8.82% | 7.70% | 7.09% | 6.75% | 6.56% | 6.51% |
| 520 | 13.54% | 9.38% | 7.48% | 6.86% | 6.61% | 6.47% | 6.33% |

Fig. 4 and Table 1 give GEDI imputation accuracy when the number of HMM founders is varied between 3 and 15 and the number of training haplotypes is varied between 30 and 520. Accuracy improves significantly when using reference panels larger than the commonly used Hapmap panels, particularly in conjunction with increasing the number of HMM founders. For example, compared to the GEDI settings used in [4] (120 training haplotypes and 7 founders), increasing the number of training haplotypes to 520 and the number of founders to 15 yields an accuracy gain of over 2.5%.

Although the accuracy gained by using a larger number of HMM founders comes at the cost of increased imputation time, the latter remains practical for up to 15 founders, above which accuracy gains become very small. Indeed, as shown in Fig. 5, GEDI optimizations such as the trie speed-up described in Section 3.1 lead to sub-cubic runtime growth within the tested range of HMM founders, allowing users to better control the tradeoff between imputation speed and accuracy.

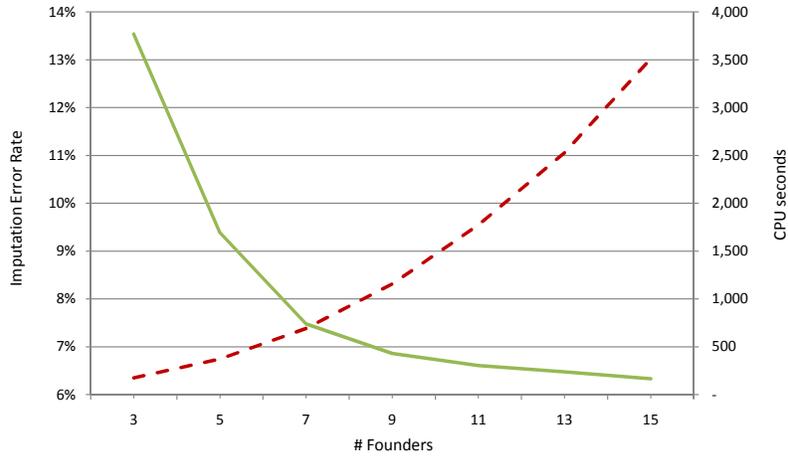

Figure 5: GEDI imputation error rate and runtime for varying number of founders (IMAGE chr. 22 dataset, 520 training haplotypes).

GEDI is also able to exploit pedigree information when available. For genotype data of related individuals, imputation probabilities (and log-likelihood ratios) are computed jointly over parents-child trios, using an extended version of the forward-backward algorithm in

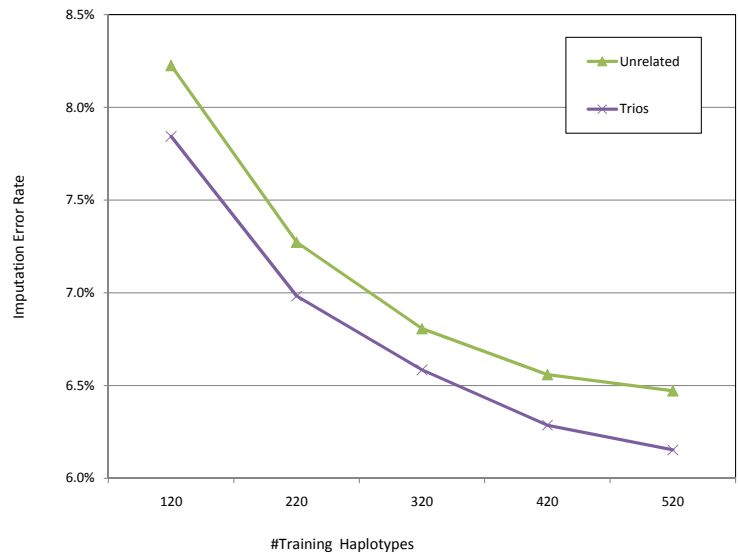

Figure 6: Effect of using pedigree information during imputation (IMAGE chr. 22 dataset, 13 HMM founders).

Table 2: Comparison of two GEDI imputation flows on a version of the IMAGE chr. 22 dataset generated by randomly inserting 1% errors and 1% missing data (520 training haplotypes).

| GEDI flow | 7 Founders | | 13 Founders | |
|---|---|---|---|---|
| | Error Rate | CPU sec. | Error Rate | CPU sec. |
| IMP | 8.17% | 410 | 7.20% | 1,637 |
| EDC+MDR+IMP | 8.07% | 4,153 | 6.91% | 15,937 |

Section 3 (see [9] for details). Fig. 6) compares the imputation error achieved by running GEDI with 13 HMM founders on the IMAGE dataset under two scenarios: (a) treating the 2502 test individuals as unrelated (as we have done in all previous experiments), and (b) analyzing them as 834 parents-child trios. Performing trio-based imputation reduces error rate by 0.22-0.44%, depending on the number of haplotypes used for training the model, pointing out to the value of using pedigree information.

Finally, we conducted experiments to assess the value of performing genotype error correction and missing data recovery prior to imputation. We generated a version of the IMAGE chr. 22 dataset generated by randomly inserting 1% errors and 1% missing data at typed SNP loci, and then ran two different analysis flows provided by GEDI:

- In the first flow, referred to as IMP, genotypes at untyped SNP (the same as those used in previous experiments) were imputed based on the genotype data at typed SNPs and HMM models trained using 520 reference haplotypes.

- In the second flow, referred to as EDC+MDR+IMP, we first trained an HMM model over typed SNPs using the 520 reference haplotypes together with haplotypes inferred by phasing all test genotypes. This model was next used to run GEDI's error detection and correction and missing data recovery functions, replacing every SNP genotype $g_i$ for which the likelihood ratio $\max_x P_M(\mathbf{g}[g_i \leftarrow x])/P_M(\mathbf{g})$ is greater than $10^3$, respectively every missing SNP genotype $g_i$, by $\mathrm{argmax}_x P_M(\mathbf{g}[g_i \leftarrow x])$. Finally, imputation was performed as in the IMP flow, but based on the modified genotype data for typed SNPs rather than the original genotypes.

Table 2 gives the error rate and runtime for running the two flows with 7, respectively 13 HMM founders. Performing the EDC+MDR+IMP flow improves accuracy over direct imputation in both cases, by almost 0.3% in the case of 13 founders. While EDC+MDR+IMP requires about 10× more time for the IMAGE dataset used in our experiment, the runtime increase should be much smaller for typical GWAS datasets, for which the number of typed loci is typically smaller than that of untyped loci. Indeed, for such datasets imputation time (which grows linearly with the number of untyped loci) is likely to dominate the time needed for performing error detection and correction and missing data recovery (which is proportional to the number of typed loci).

While the accuracy gains obtained by using pedigree information or performing the EDC+MDR+IMP flow are small, they can translate in non-negligible cost savings. In-deed, as noted by [7], each 1% gain in imputation accuracy translates into a 5-10% reduction in the sample size needed to achieve a desired statistical power level.

# Acknowledgments

This work was supported in part by National Science Foundation awards IIS-0546457, IIS-0916948, and DBI-0543365. The IMAGE dataset used for the experiments described in this report was obtained through accession number phs000016.v1.p1 from the database of Genotype and Phenotype (dbGaP) found at http://www.ncbi.nlm.nih.gov/gap. Genotyping of IMAGE samples was provided through the Genetic Association Information Network (GAIN).

# References

[1] A.S. Allen and G.A. Satten. A novel haplotype-sharing approach for genome-wide case-control association studies implicates the calpastatin gene in Parkinson's disease. *Genet. Epidemiol.,* Epub ahead of print, 2009.

[2] B. L. Browning and S. R. Browning. Haplotypic analysis of Wellcome Trust case control consortium data. *Human Genetics*, 123(3):273–280, 2008.

[3] F. Dudbridge. Pedigree disequilibrium tests for multilocus haplotypes. *Genet. Epidemiol.*, 25:115–221, 2003.

[4] GAIN Imputation Working Group. In preparation. 2009.

[5] A. Gusev, I.I. Măndoiu, and B. Paşaniuc. Highly scalable genotype phasing by entropy minimization. *IEEE/ACM Trans. Comp. Biol. Bioinf.*, 5(2):252–261, 2008.

[6] L.A. Hindorff, P. Sethupathy, H.A. Junkins, E.M. Ramos, J.P. Mehta, F.S. Collins, and T.A. Manolio. Potential etiologic and functional implications of genome-wide association loci for human diseases and traits. *Proceedings of the National Academy of Sciences*, 106(23):9362–9367, 2009.

[7] L. Huang, C. Wang, and N.A. Rosenberg. The relationship between imputation error and statistical power in genetic association studies in diverse populations. *Am. J. Hum. Genet., in press*, 2009.

[8] International HapMap Consortium. A second generation human haplotype map of over 3.1 million SNPs. *Nature*, 449(7164):851–861, 2007.

[9] J. Kennedy, I.I. Măndoiu, and B. Paşaniuc. Genotype error detection using hidden markov models of haplotype diversity. *J. Comput. Biol.*, 15(9):1155–1171, 2008.

[10] M. Knapp and T. Becker. Impact of genotyping errors on type I error rate of the haplotype-sharing transmission/disequilibrium test (HS-TDT). *Am. J. Hum. Genet.*, 74:589–591, 2004.

[11] Y. Li and G. R. Abecasis. Mach 1.0: Rapid haplotype reconstruction and missing genotype inference. *Am. J. Hum. Genet.*, 79:2290, 2006.

[12] D.Y. Lin, Y. Hu, and B.E. Huang. Simple and efficient analysis of disease association with missing genotype data. *Am. J. Hum. Genet.*, 83(4):535–539, 2008.

[13] J. Marchini, B. Howie, S. Myers, G. McVean, and P. Donnelly. A new multipoint method for genome-wide association studies by imputation of genotypes. *Nat. Genet.*, 39:906–913, 2007.

[14] Neale *et al.* Genome-wide association scan of attention deficit hyperactivity disorder. *Am. J. Med. Genet. Part B*, 147b(8):1337–1344, 2008.

[15] P. Scheet and M. Stephens. A fast and flexible statistical model for large-scale population genotype data: applications to inferring missing genotypes and haplotypic phase. *Am. J. Hum. Genet.*, 78(4):629–644, 2006.

[16] N. Zaitlen, H. Kang, M. Feolo, S. T. Sherry, E. Halperin, and E. Eskin. Inference and analysis of haplotypes from combined genotyping studies deposited in dbSNP. *Genome Research*, 15:1595–1600, 2005.

[17] Zeggini *et al.* Meta-analysis of genome-wide association data and large-scale replication identifies additional susceptibility loci for type 2 diabetes. *Nat. Genet.*, 40(5):638–645, 2008.